\documentclass[referee]{raa}            % referee version: for submission

\usepackage{graphicx,times}             %for PS/EPS graphics inclusion, new

\begin{document}

   \title{Physical parameters and orbital period variation of a newly discovered cataclysmic variable GSC 4560-02157
%\footnotetext{$*$ Supported by the National Natural Science Foundation of China.}
}
%   \subtitle{I. Place Your Subtitle Here}

   \volnopage{Vol.0 (200x) No.0, 000--000}      %%preserved for Editor. DOn't remove!
   \setcounter{page}{1}          %%starting page, preserved for Editor. DOn't remove!

   \author{Zhong-tao Han
      \inst{1,2,3}
   \and Sheng-bang Qian
      \inst{1,2,3}
   \and Irina Voloshina
      \inst{4}
   \and Vladimir G.Metlov
      \inst{4}
   \and Li-ying Zhu
      \inst{1,2,3}
   \and Lin-jia Li
      \inst{1,2}
   }

   \institute{Yunnan Observatories, Chinese Academy of Sciences (CAS), P. O. Box 110, 650216 Kunming, China; {\it zhongtaohan@ynao.ac.cn}\\
        \and
             Key Laboratory of the Structure and Evolution of Celestial Objects, Chinese Academy of Sciences, P. O. Box 110, 650216 Kunming, China\\
        \and
             University of Chinese Academy of Sciences, Yuquan Road 19\#, Sijingshang Block, 100049 Beijing, China\\
        \and
             Sternberg Astronomical Institute, Moscow State University, Universitetskij prospect 13, Moscow 119992, Russia\\
   }

   \date{Received~~2009 month day; accepted~~2009~~month day}

\abstract{GSC 4560-02157 is a new eclipsing cataclysmic variable with an orbital period of $0.265359$ days. By using the published $V-$ and $R-$band data together with our observations, we discovered that the $O-C$ curve of GSC 4560-02157 may shows a cyclic variation with the period of $3.51$ years and an amplitude of $1.40$ min. If this variation is caused by a light travel-time effect via the existence of a third body, its mass can be derived as $M_{3}sini'\approx91.08M_{Jup}$, it should be a low-mass star.
In addition, several physical parameters were measured. The colour of the secondary star was determined as $V-R=0.77(\pm0.03)$ which corresponds to a spectral type of K2-3. The secondary star's mass was estimated as $M_{2}=0.73(\pm0.02)M_{\odot}$ by combing the derived $V-R$ value around phase 0 with the assumption that it obeys the mass-luminosity relation of the main sequence stars. This mass is consistent with the mass$-$period relation of CV donor stars. For the white dwarf, the eclipse durations and contacts of the white dwarf yield an upper limit of the white dwarf's radius corresponding to a lower limit mass of $M_{1}\approx0.501M_{\odot}$. The overestimated radius and previously published spectral data indicate that the boundary layer may has a very high temperature.
}

   \authorrunning{Zhong-tao Han et al. }            %author_head in even pages
   \titlerunning{A new cataclysmic variable GSC 4560-02157 }  % title_head in odd pages

   \maketitle
%% The author head (on even pages) and the title head (on odd pages) will be
%% automatically extracted from \author{} and \title{}. Whenever the title is too long,
%% you will be asked to supply a shorter one by inserting either \authorrunning{} or
%% \titlerunning{} before \maketitle. Anyway, you can specify your own heads.
%%
%%
%% Note: In the following text body of your manuscript, please note several differences from
%%       other major journals:
%% (1) \subsection{Please Capitalize the First Letter of Each Notional Word in Subsection Title}
%% (2) Please Capitalize the First Letter of Each Notional Word in all tables' captions

%
%________________________________________________ sections below
%
\section{Introduction}           %% first-level sections will be auto-capitalized
\label{sect:intro}
Cataclysmic variables (CVs) are close binaries containing a low-mass main sequence star filling its Roche lobe and a white dwarf (Warner~\cite{wr1995}). As the mass flows toward the white dwarf, an accretion disc around the white dwarf was formed. When a white dwarf is in an eclipsing CV, the eclipse times can be used to probe the companion objects and its orbital evolution. The orbital periods of several CVs have been exhibited cyclic changes and were interpreted as the presence of the circumbinary companions such as V2051 Oph (Qian et al.~\cite{Qian2015}), Z Cha (Dai et al.~\cite{Dai2009}) and OY Car (Han et al.~\cite{Han2015}) etc. The discovery of a companion object orbiting a CV have important implications for understanding of extrasolar planet formation and evolution. Since these binary systems form as a result of a common envelope (CE) phase (Andronov et al.~\cite{An2003}).
In addition, the eclipsing CVs also provide a good opportunity to measure the fundamental parameters and give some constraints on system structures.

GSC 4560-02157 is a newly discovered eclipsing CV (Khruslov et al.~\cite{Kh2014}). This CV was found to be a variable star from the Northern Sky Variability Survey Database (Wozniak et al.~\cite{Woz2004}) and attracted a researcher (Khruslov) attention in 2005.
Khruslov et al.~\cite{Kh2014} reported spectroscopic and many photometric observations, the results shown that GSC 4560-02157 is an eclipsing binay with a long orbital period of about 6.37 hours. By the photometric observations, they found that the out-of-eclipse brightness demonstrates sensible changes but the primary eclipse was relatively stable. Moreover, its eclipse profile and spectral feature are very similar to another long period eclipsing CV GY Cnc. Lastly, these authors confirmed that GSC 4560-02157 is an eclipsing CV. In present paper, by using the data published by Khruslov et al.~\cite{Kh2014}, we obtained some basic system parameters and mid-eclipse times. Combining these mid-eclipse times with our latest data, the linear ephemeris was revised and a cyclic variable in $O-C$ diagram was found.

%% Authors can give a citation as 'Michel et al. 1992'.
%% You may also use \cite, \citep and \citet for citation, and use Table~1 or Figure~1
%% and so forth. Using \ref and \label for cross-references of Tables/Figures
%% is a good way in adjusting/adding/removing text, tables or figures.

\section{Observations and Data reduction}
\label{sect:Obs}
GSC 4560-02157 was observed with CCD photometer on 50-cm (Apogee Alta U8300 with 528 x 512 pixels) telescope of Sternberg Astronomical Institute Crimean Station on 22 and 28, April 2016, respectively. During the observations, $R-$band were used. Aperture photometry was performed with MAXIM DL package. All mid-eclipse times (including the data published by Khruslov et al.~\cite{Kh2014}) were determined by using a parabolic fitting method because the light variations around those minima are near-symmetrical. All available $CCD$ mid-eclipse times of GSC 4560-02157 are listed in Table $1$. Two eclipse profiles from our observation are displayed in Fig.1. For comparison, another two published eclipsing light curves are shown in Fig.2. These light curves exhibit the primary eclipse and an similar feature with Algol-type eclipsing binary. However, the highly variables outside of eclipse indicated that it is more likely the IP Peg-like or GY Cnc-like type. In addition, the secondary eclipse can be seen clearly in the light curve and the brightness variations on a variety of scales are superimposed on it.
\begin{figure}
\begin{center}
\includegraphics[width=\columnwidth]{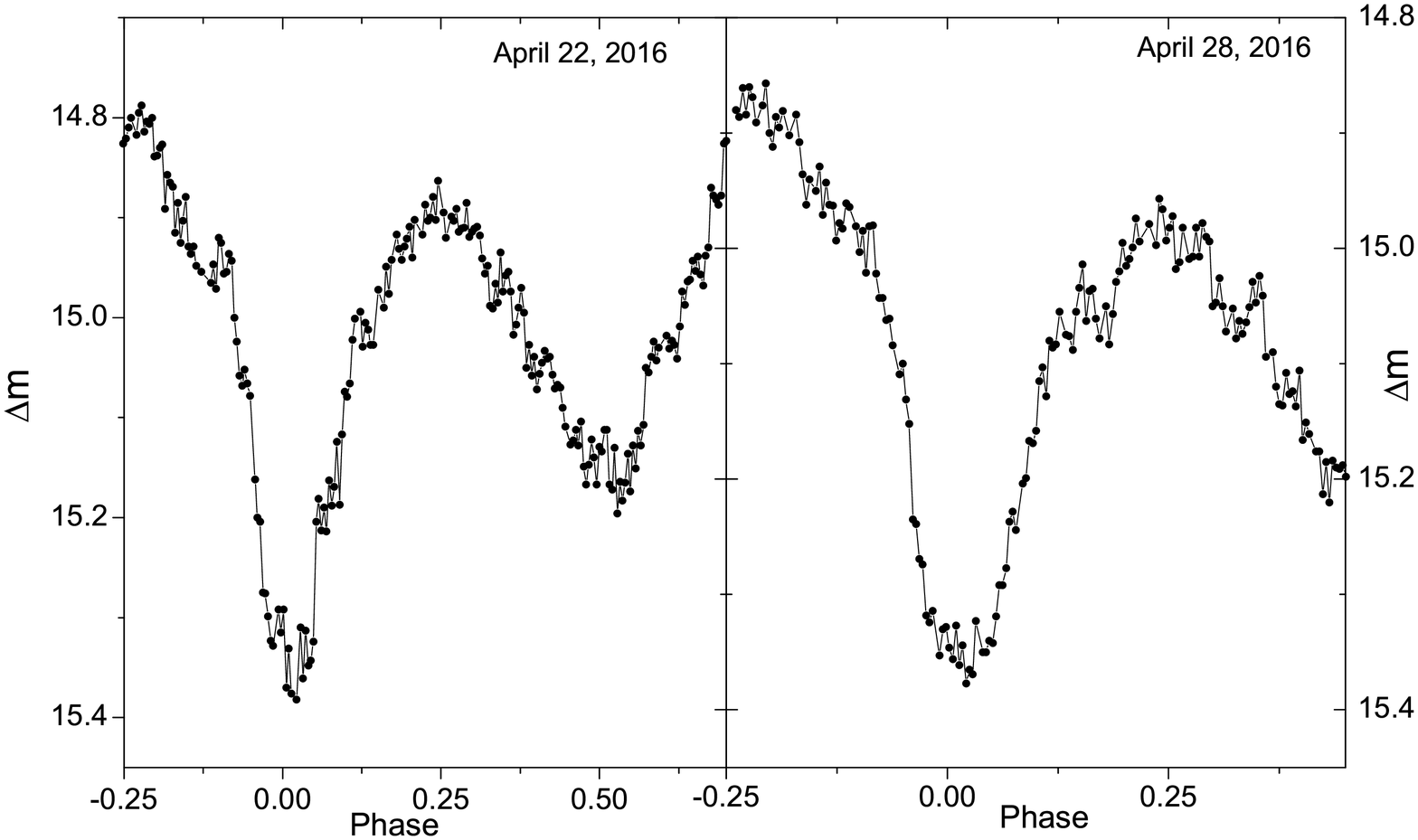}
\caption{Two eclipsing profiles of GSC 4560-02157 are shown. The data were observed by using the 50cm telescope in Sternberg Astronomical Institute Crimean Station in 2016.}
\label{fig:figure1.eps}
\end{center}
\end{figure}
\begin{figure}
\begin{center}
\includegraphics[width=\columnwidth]{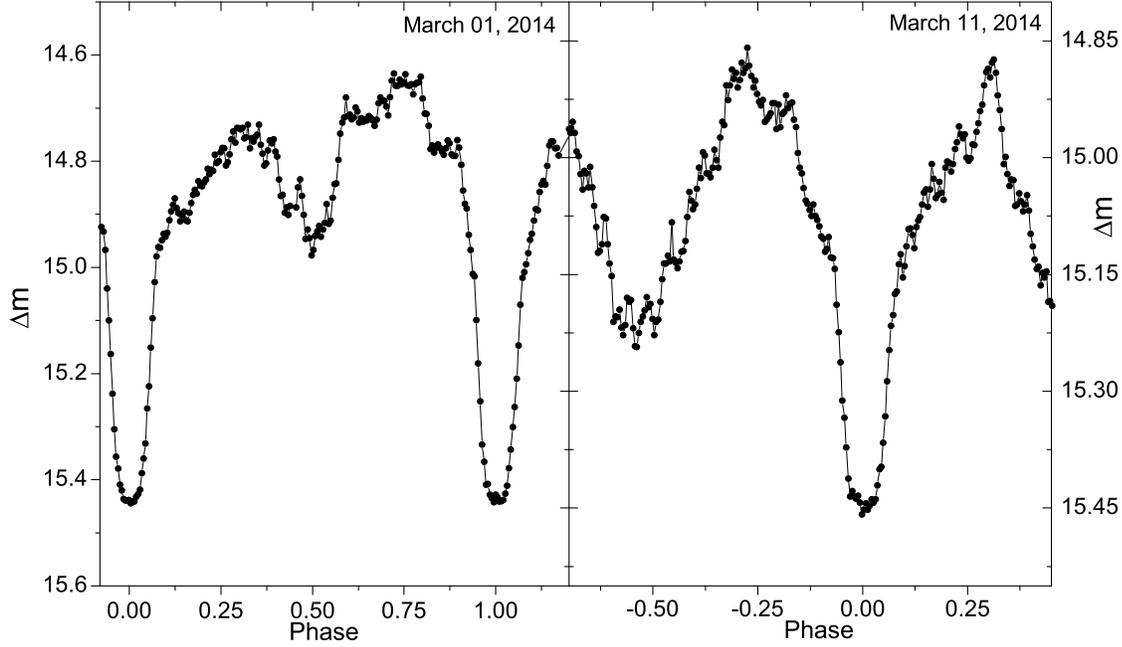}
\caption{The data from Khruslov et al.~\cite{Kh2014}. These light curves exhibit a strong change of the out-of-eclipse brightness.}
\label{fig:figure2.eps}
\end{center}
\end{figure}

\section{Results}
\subsection{Revised ephemeris}
Since its discovery, the changes of the orbital period have never analysed. By adopting the photometric data from Khruslov et al.~\cite{Kh2014}, a total of 16 minima were obtained. Besides, we also observed it on 22 and 28, April 2016 and 2 mid-eclipse times were measured. All data are listed in Table 1. The $O-C$ values of all minima were computed with the following linear ephemeris:
\begin{equation}
Min.I=HJD2456594.0637+0.265359\times{E},
\end{equation}
where HJD2456594.0637 is the initial epoch from the first column in Table 1, 0.209937316 is the orbital period from Khruslov et al.~\cite{Kh2014}, $E$ is the cycle number. The newest $O-C$ diagram is shown in Fig.3. Overall, the $O-C$ curve shows a possible cyclic variation, and then a sinusoidal function can be considered to describe it. The result of the best-fitting is as follows:
\begin{equation}
\begin{array}{lll}
O-C=&0.00112(\pm0.00042)+1.051(\pm0.026)\times10^{-7}\times{E}\\
          &+0.00097(\pm0.00026)\sin[0^{\circ}.0745(\pm0^{\circ}.0006)\times{E}-82^{\circ}.29(\pm0^{\circ}.01)],
\end{array}
\end{equation}
with $\chi^{2}=0.00032$ (the sum of square residuals). Clearly, the sinusoid model displayed
in Fig.3 is a good description. This equation indicates a cyclic change with an amplitude of $1.40$ min and a period of $3.51$ yrs. In Fig.3, the solid line in the upper panel refers to the best-fitting sine function, the residuals are plotted in the bottom panel.
\begin{figure}
\begin{center}
\includegraphics[width=\columnwidth]{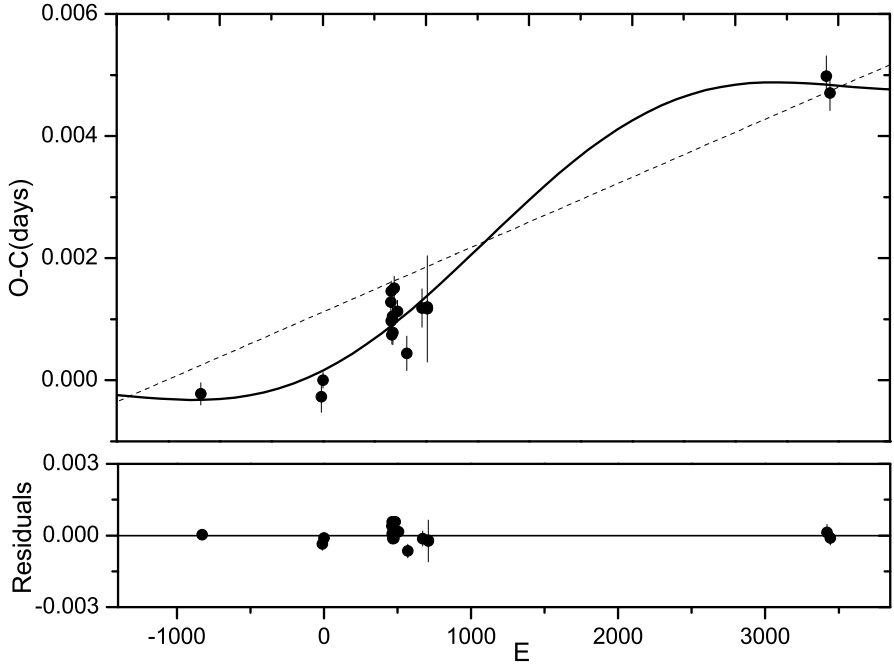}
\caption{$O-C$ diagrams of GSC 4560-02157. The solid line in the upper panel refers to a cyclic change. The dashed line represents a revised linear ephemeris. Both these were removed, the residuals are plotted in the bottom panel.}
\label{fig:figure3.eps}
\end{center}
\end{figure}

\begin{table*}
\centering
\caption{New CCD mid-eclipse times of GSC 4560-02157. References: (1). Khruslov et al. 2014; (2). Our data.}
\label{tab:table1}
 \begin{center}
 \small
 \begin{tabular}{lllllll}
   \hline
   \hline
Min.(HJD)        &Err       &E        &O-C       &Fil.     &Durations   &References\\\hline
2456374.34671    &0.00018   &-828    &-0.00022   &R        &0.0356      &(1)\\
2456591.14506    &0.00025   &-10     &-0.00027   &R        &0.0364      &(1)\\
2456594.06374    &0.00013   &0        &0.00000   &R        &0.0318      &(1)\\
2456716.39552    &0.00020   &461      &0.00128   &R        &0.0305      &(1)\\
2456717.19129    &0.00012   &464      &0.00097   &R        &0.0318      &(1)\\
2456717.45713    &0.00016   &465      &0.00146   &R        &0.0318      &(1)\\
2456718.25249    &0.00015   &468      &0.00075   &R        &0.0319      &(1)\\
2456719.31423    &0.00018   &472      &0.00105   &R        &0.0305      &(1)\\
2456720.37540    &0.00020   &476      &0.00078   &R        &0.0342      &(1)\\
2456722.49900    &0.00019   &484      &0.00151   &R        &0.0354      &(1)\\
2456728.33652    &0.00018   &506      &0.00113   &R        &0.0332      &(1)\\
2456745.31881    &0.00031   &570      &0.00044   &V        &0.0321      &(1)\\
2456772.38606    &0.00011   &672      &0.00107   &R        &0.0341      &(1)\\
2456772.38629    &0.00028   &672      &0.00130   &V        &0.0334      &(1)\\
2456782.20444    &0.00034   &709      &0.00120   &R        &0.0342      &(1)\\
2456782.20447    &0.00087   &709      &0.00117   &V        &0.0304      &(1)\\
2457501.33114    &0.00033   &3419     &0.00498   &R        &0.0343      &(2)\\
2457507.43412    &0.00029   &3442     &0.00471   &R        &0.0360      &(2)\\\hline
\end{tabular}
\end{center}
\end{table*}

\subsection{Physical parameters}
\label{sect:data}
In this section, some physical parameters will be calculated and discussed. Many $V-$ and $R-$band light curves of GSC 4560-02157 were published by Khruslov et al.~\cite{Kh2014}. However, only one system parameter was obtained, i.e. orbital period ($0.265359$ d). In principle, the $V-R$ color can be estimated by adopting $V-$ and $R-$band light curves. In the calculation process, the color index of primary minimum was used to determine the spectral type of secondary star owing to primary star and other light sources (including accretion disc and bright spot) were almost entirely eclipsed by secondary star. Due to the data from Khruslov et al.~\cite{Kh2014} were performed by using differential photometry method, the $V-R$ values of comparison star are needed. The following formula,
\begin{equation}
\Delta{V}-\Delta{R}=(V-R)-(V-R)_{C}
\end{equation}
was used to calculate the $V-R$ color. $(\Delta{V}-\Delta{R})$ and $(V-R)_{C}$ in this formula are the color index from the differential photometry method and the comparison star, respectively.
The comparison star is GSC 4560-01221 in GSC2.3 catalog (Khruslov et al.~\cite{Kh2014}; Lasker et al.~\cite{Las2008} ), its magnitudes for $V-$ and $R-$band are $14.33$ and $14.11$ mag, respectively. The values of $(\Delta{V}-\Delta{R})$ are computed by using $V-$ and $R-$band light curves offered by Khruslov et al.~\cite{Kh2014} on 25, April and 05, May 2014. Lastly, the $V-R$ value around phase 0 was derived as $0.77(\pm0.03)$. Since the secondary star of GSC 4560-02157 may be a main sequence star, according to the MK spectral classes (Cox A.~\cite{Cox2000}, Table 15.7), the spectral type could be roughly estimated as K2$-$K3 even later, corresponding to $T_{2}\leq4600-4800$ K.
Cox A. ~\cite{Cox2000} also gave the calibrated physical parameters for stars of the various spectral classes (Cox A.~\cite{Cox2000}, Table 15.8), the mass-luminosity relation gives a mass of $M_{2}\approx0.73(\pm0.02)M_{\odot}$ for K2$-$3 type. For CVs, their mass-radius relationship will be roughly (Knigge~\cite{Kni2011} )
\begin{equation}
R_{2}/R_{\odot}=f(M_{2}/M_{\odot})^{\alpha}
\end{equation}
with $f\simeq\alpha\simeq1$. $M_{2}\approx0.73M_{\odot}$ corresponds to $R_{2}\approx0.73R_{\odot}$. For comparison, we put this value in the mass$-$period relation of CV donor stars. As shown in Fig.4, the location of this donor (the asterisk) is compatible with the overall trend.

\begin{figure}
\begin{center}
\includegraphics[width=\columnwidth]{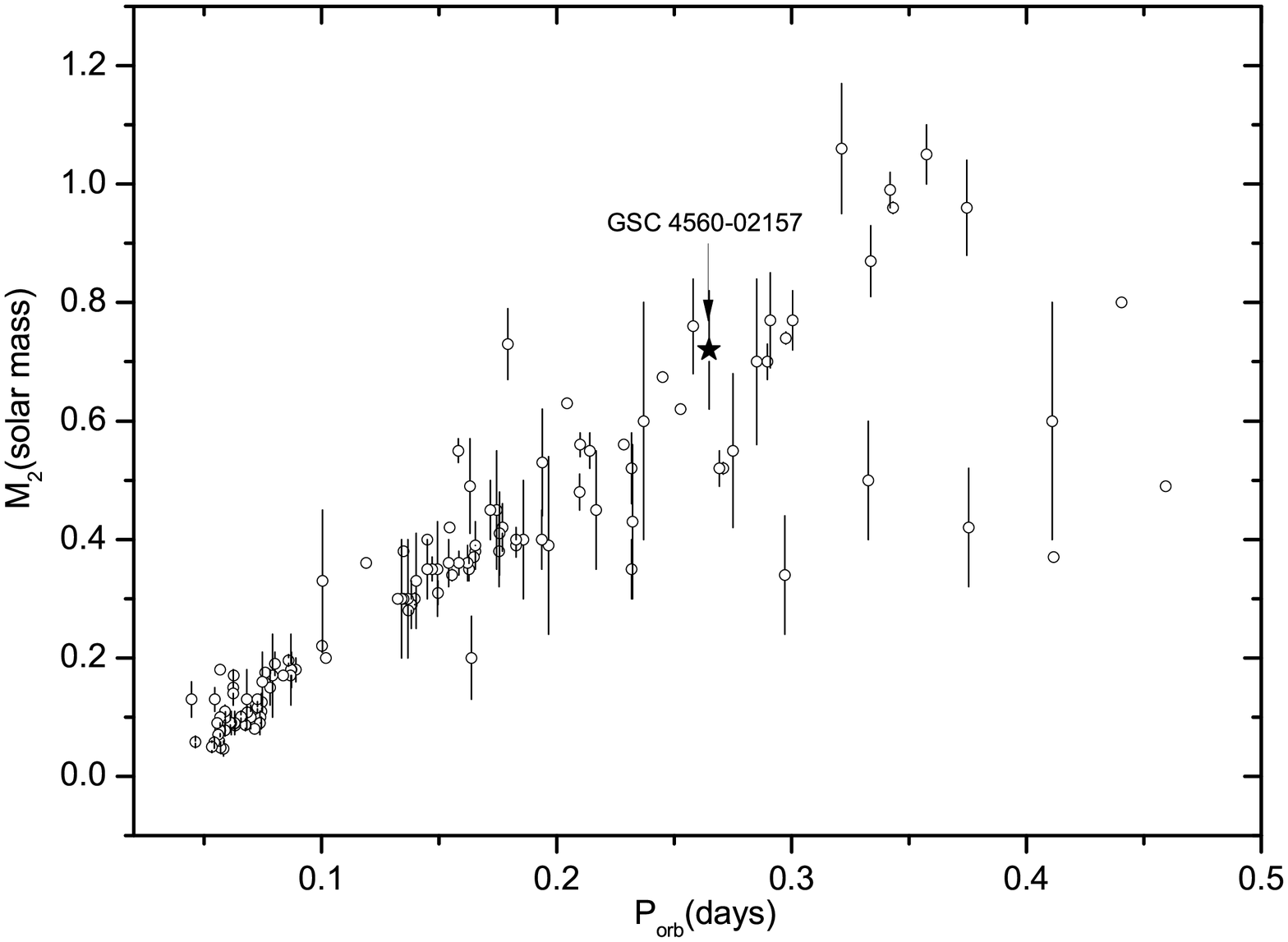}
\caption{Mass-period relation of CV donor stars. Data from the latest version (update RKcat7.23, 2015) of the Ritter \& Kolb~\cite{Rit2003} catalogue. The mark with an asterisk denoted the location of GSC 4560-02157.}
\label{fig:figure4.eps}
\end{center}
\end{figure}
\begin{figure}
\begin{center}
\includegraphics[width=\columnwidth]{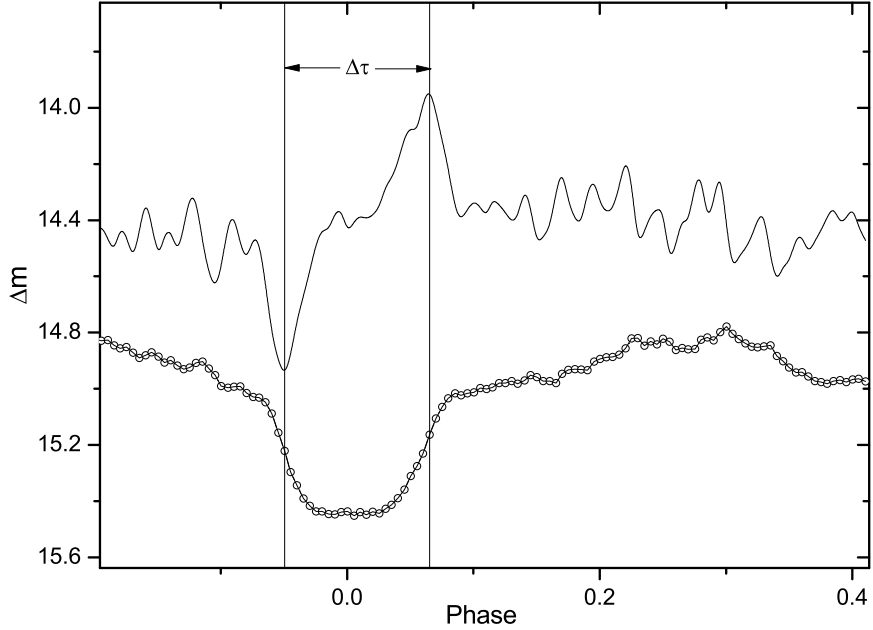}
\caption{Example of determining eclipse durations. The interval of the solid lines represents the the durations.}
\label{fig:figure5.eps}
\end{center}
\end{figure}
\begin{figure}
\begin{center}
\includegraphics[width=\columnwidth]{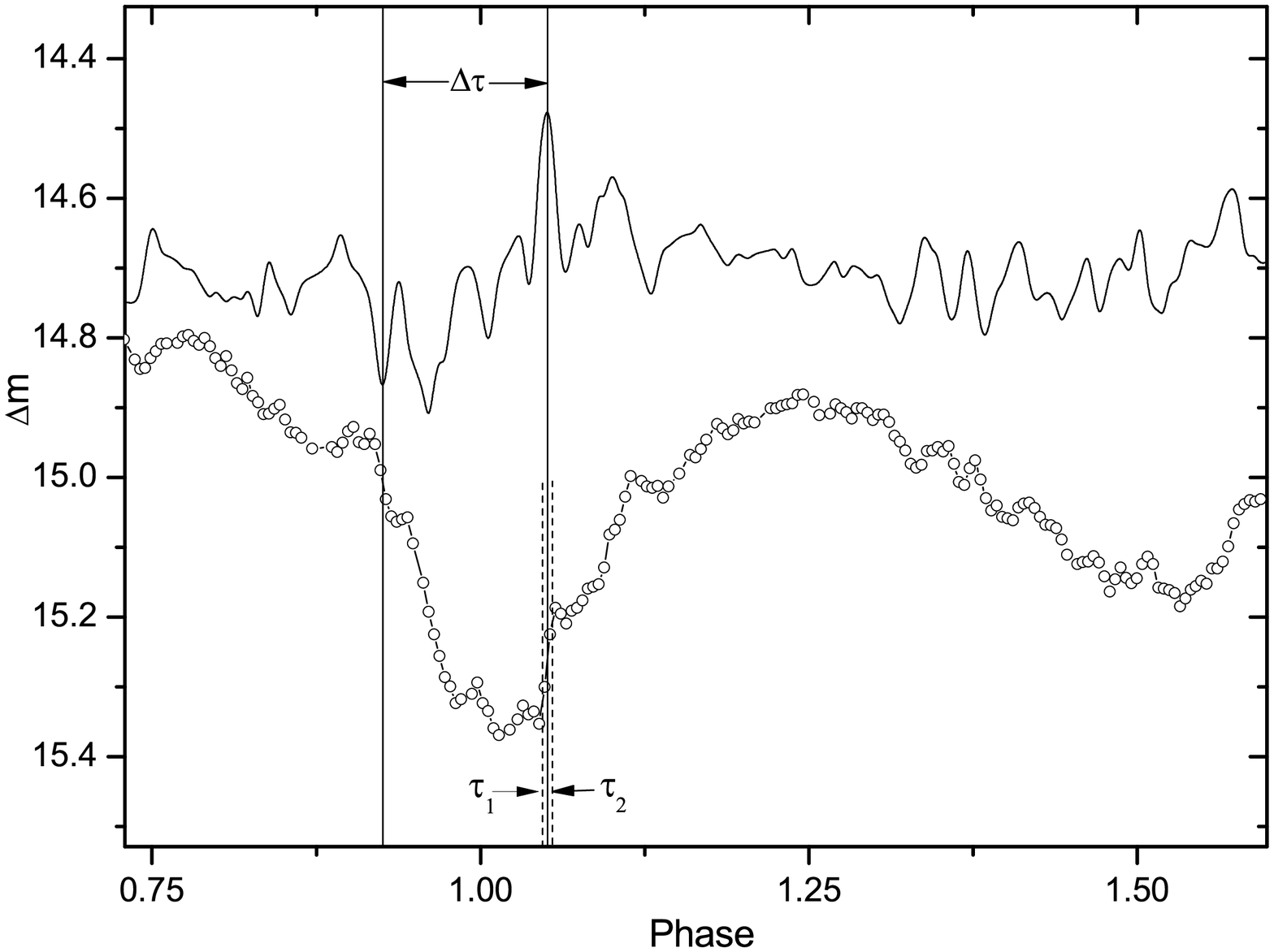}
\caption{A sample of measuring the contacts of the white dwarf. The two dashed lines refer to the start and end times of the white dwarf egress, respectively.}
\label{fig:figure6.eps}
\end{center}
\end{figure}
For the white dwarf, theoretically, its parameters can be derived from eclipsing CVs if eclipse contacts of the white dwarf can be measured precisely. However, the radial velocity of two components are unknown. To resolve this problem further, the eclipse duration and contact of the white dwarf can be considered to determine radii of stars. For eclipse durations, the eclipse profiles from  Khruslov et al.~\cite{Kh2014} are almost symmetric and have a relatively flat bottom, suggesting that the white dwarf should be the center of the bright source. Therefore, the durations can be determined by locating the extreme value of the eclipse curve derivative occurred times. This method was described by Wood et al.~\cite{wo1985}. First, the observed light curves were smoothed with a median filter and differentiated numerically. Second, the derivative curves were used to located the extreme points. This measurement procedure was illustrated in Fig.5. All durations were measured and listed in sixth column of Table 1. However, the eclipse contacts of the white dwarf are very difficult to determine because the boundary layer and a hot inner-disc region around the white dwarf. Fortunately, a well-defined egress profile of the white dwarf was observed (see Fig.6). This eclipse profile has two distinct steps corresponding to the occultation of the white dwarf and bright spot. These features are similar to some typical eclipsing dwarf novae such as OY Car (Wood et al.~\cite{Wo1989}) and Z Cha (Wood et al.~\cite{Wo1986}) etc. As shown in Fig.6, $\Delta\tau$ refers to the duration, $\tau_{1}$ and $\tau_{2}$ is the start and end of the white dwarf egress, respectively. It should be noted that the white dwarf's ingress is not used to measure the start and end times of the white dwarf ingress owing to the ingress times of the white dwarf and bright spot may be closely simultaneous. The radii of both components were derived by using the following equations:
\begin{equation}
R_{1}=\frac{v_{1}+v_{2}}{2}(\tau_2-\tau_1)
\end{equation}
and
\begin{equation}
R_{2}=\frac{v_{1}+v_{2}}{2}\Delta\overline{\tau}
\end{equation}
where $v_{1}$ and $v_{2}$ refer to the radial velocity of primary and secondary star, and $\Delta\overline{\tau}$ is the average value of all durations in Table 1. Combining equation (5) with (6) immediately yields the white dwarf's radius as $R_{1}\approx0.0234R_{\odot}$. During the calculation, $\tau_2-\tau_1\approx1.57$ min was used.

\begin{figure}
\begin{center}
\includegraphics[width=\columnwidth]{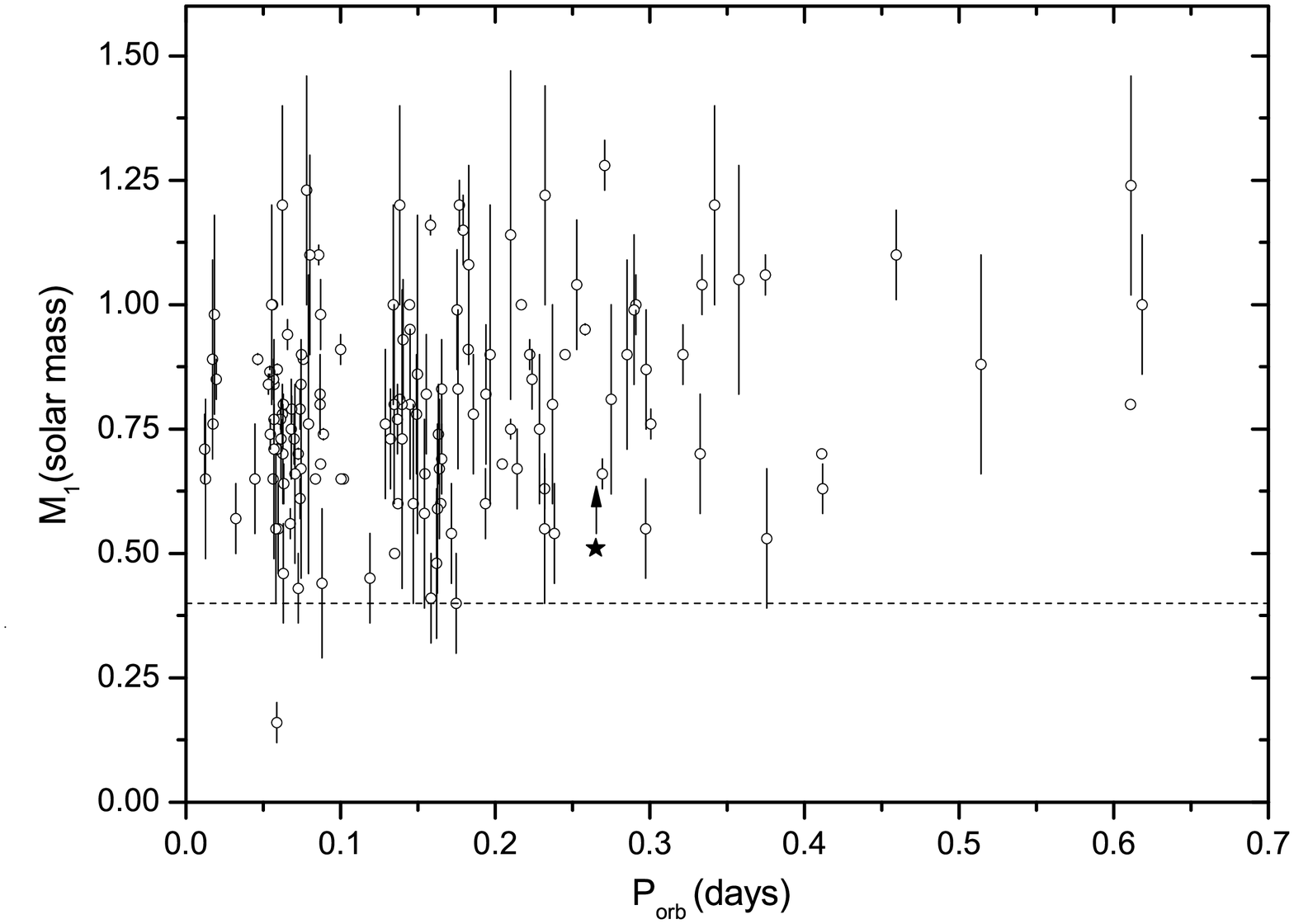}
\caption{Mass-period relation of CV primary stars. Data from the latest version (update RKcat7.23, 2015) of the Ritter \& Kolb~\cite{Rit2003} catalogue. The position of the white dwarf of GSC 4560-02157 in relation is the mark with an asterisk.}
\label{fig:figure7.eps}
\end{center}
\end{figure}
If we used the mass-radius relation of white dwarfs to estimate its mass (Michael~\cite{Mic1972}), $0.0234R_{\odot}$ corresponds to $0.225M_{\odot}$. This value in mass$-$period diagram is much lower than the all CVs with $P_{orb}>0.2$ days. For the evolutionary perspective, this value is also too small. There are several reasons for overestimating white dwarf's radius. One is that the white dwarf may has a 'thick' boundary layer or atmosphere due to its high accretion (see Section 4). Another reason is that supposing orbital inclination equals $90^{\circ}$ during the calculation.
Parsons et al.~\cite{Pa2010} have been accurately measured the white dwarf's mass and radius of a pre-CV NN Ser independently of any mass-radius relation. They gave the values as $R_{1}=0.0211R_{\odot}$ and $M_{1}=0.535M_{\odot}$, and also calculated the surface gravity of the white dwarf as $\log{g}=7.74$. This value may be a reasonable approximation for the white dwarf of GSC 4560-02157. Adopting $\log{g}=7.74$ for GSC 4560-02157, the mass of the white dwarf can be estimated as $M_{1}\approx0.501M_{\odot}$. However, the derived white dwarf'mass ($0.501M_{\odot}$) is only a lower limit corresponding to an upper limit of its radius. Fig.7 shows the position of this lower limit value in the mass$-$period relation of CV primary stars.

\section{Discussion and Conclusion}
\label{sect:analysis}
Monitoring the mid-eclipse times of CVs provides insight into the orbital evolution. As noted above, the $O-C$ curve of GSC 4560-02157 may shows a cyclic change with a period of $3.51$ yr. The likely    explanations for this cyclic variation are either a solar-like magnetic activity in the secondary star (Applegate~\cite{Ap1992} ) or a third body in the system. Most CVs with period modulations can be found from Pilar{\v c}{\'i}k et al.~\cite{Pil2012}. If the periodic change of the $O-C$ diagram is caused by a companion orbiting GSC 4560-02157, the mass function and the mass of this companion were computed by using the following equation:
\begin{equation}
f(m)=\frac{4\pi^{2}}{GP_{3}^{2}}(a'_{12}sini')^{3}=\frac{(M_{3}sini')^{3}}{(M_{1}+M_{2}+M_{3})^{2}},
\end{equation}
where $G$ is the gravitational constant, $P_3$ is the period of the $O-C$ oscillation, and $a'_{12}sini'$ can be determined by
\begin{equation}
a'_{12}sini'=K_{3}\times c ,
\end{equation}
$K_3$ is the semi-amplitude of the $O-C$ oscillation, i.e. $a'_{12}sini'=0.168(\pm0.03)$ AU. By using the parameters from Section 3.2, the mass function and the mass of third body were derived as $f(m)=3.85(\pm2.4)\times10^{-4}M_{\odot}$ and $M_{3}sini'=0.087(\pm0.032)M_{\odot}\approx91.08M_{Jup}>0.075M_{\odot}$, respectively. Clearly, this is a low-mass star. The parameters of the tertiary component are displayed in Table 2. However, a large error range of this companion mass means it may be a brown dwarf with a small probability. As expected, no long-term change in $O-C$ curve was discovered. The main reason is that the time span of the data is less than 3 years.
\begin{table}[h]
\center
 \normalsize
 \caption{The orbital parameters of the third body in GSC 4560-02157.}
\begin{tabular}{lll}
\hline\hline
Parameters              &    Value and uncertainty    &  Unit \\
\hline
Period($P_{3}$)         &    3.51$\pm$ 0.01           &  yrs \\
Eccentricity($e_{3}$)   &    0                        &\\
Amplitude($K_{3}$)      &    1.40$\pm$0.02            & min \\
$a_{3}(i'=90^{\circ})$  &    2.81$\pm$ 0.52           &  AU \\
$M_{3}sini'$            &    91.08$\pm$34.02          &  $M_{Jup}$ \\
$f$($m$)                &    3.85$\pm$(2.4)$\times10^{-4}$ & $M_{\odot}$
\\
\hline
\end{tabular}
\end{table}

Additionally, combining the published $V-$ and $R-$band data with our observations, several physical parameters were obtained. The derived mass of donor star is about $0.73(\pm0.02)M_{\odot}$, consistent with mass$-$period relation of CV donor stars. As noted above, the measured radius of the white dwarf are likely to be overestimated. A spectroscopic observation at the phase 0.505 published by Khruslov et al.~\cite{Kh2014} shows very obvious HeI and HeII emissions. Szkody et al.~\cite{Sz2009} pointed out that the HeII 4686 line is a strong indicator of high accretion. This means the boundary layer has a very high temperature and the disc is relatively much brighter. The inner disc may has a compact structure surrounding this white dwarf (Hoare \& Drew.~\cite{Hoa1991} ). In addition, we've assumed that the orbital inclination is $90^{\circ}$ in the computing process. Therefore, we obtained the upper limit of the white dwarf radius, corresponding to the lower limit of mass.
Certainly, much longer photometric monitoring for GSC 4560-02157 in the future are required in order to ascertain its nature such as the orbital evolution, the system structure and the possible reasons of the cyclic variations in $O-C$ curve.

\begin{acknowledgements}
This work is partly supported by Chinese Natural Science Foundation (No.11133007 and No. 11325315), the Key Research Program of the Chinese Academy of Sciences (grant No. KGZD-EW-603), the Science Foundation of Yunnan Province (No. 2012HC011 and 2013FB084), and by the Strategic Priority Research Program The Emergence of
Cosmological Structures of the Chinese Academy of Sciences (No. XDB09010202). New CCD photometric observations of GSC 4560-02157 were obtained with 50cm telescope of Sternberg Astronomical Institute Crimean Station.
\end{acknowledgements}

\label{lastpage}

\end{document}